\documentstyle{mn}
\title[EUV observations of SW UMa during superoutburst]
{A ROSAT WFC observation of SW UMa: 
the EUV behaviour of dwarf novae in superoutburst explained}

\author[M.\,R. Burleigh et~al.]
{M.\,R. Burleigh, J.\,P. Pye, S.\,W. Poulton, K.\,B. Sohl, P.\,J. Wheatley \& 
G.\,A. Wynn\\
Department of Physics and Astronomy, University 
of Leicester, University Rd., Leicester, LE1 7RH \\
}
\date{March 27th 2001}
\newcommand{\ro}{{\sl ROSAT}}

\newcommand{\wfc}{{\sl WFC}}
\newcommand{\pspc}{{\sl PSPC}}
\newcommand{\euve}{{\sl EUVE}}
\newcommand{\EXOSAT}{{\sl EXOSAT}}
\newcommand{\ALEXIS}{{\sl ALEXIS}}

\def\la{\mathrel{\hbox{\rlap{\hbox{\lower4pt\hbox{$\sim$}}}{\raise2pt\hbox{$<$}}
}}}
\def\ga{\mathrel{\hbox{\rlap{\hbox{\lower4pt\hbox{$\sim$}}}{\raise2pt\hbox{$>$}}
}}}

\begin{document}
\maketitle

\begin{abstract}

During re-processing and analysis of the entire $\ro$ Wide Field Camera ($\wfc$) 
pointed observations database, we discovered a serendipitous, off-axis detection of the 
cataclysmic variable SW~UMa at the onset of its 1997 October superoutburst. 
Although long outbursts in this SU UMa-type system are known 
to occur every $\sim450$~days, none had ever been previously observed 
in the extreme ultra-violet (EUV) by $\ro$. 
The $\wfc$ observations began just $\approx13$~hr 
after the optical rise was detected.  
With a peak count rate of $\sim4.5$ count~s$^{-1}$ in the S1 
filter, SW~UMa was temporarily the third brightest object in the sky in 
this waveband. Over the next $\approx19$~hr the measured EUV flux dropped 
to $<2$~count~s$^{-1}$, 
while the optical brightness remained essentially static at m$_{\rm v}\sim11$. 
Similar behaviour has also been recently reported  
in the EUV light curve of the related 
SU~UMa-type binary OY~Car during superoutburst (Mauche and Raymond 2000). 
In contrast, U~Gem-type dwarf novae show no 
such early EUV dip during normal outbursts. 
Therefore, this feature may be common in superoutbursts of SU UMa-like 
systems. We expand on ideas first put forward by Osaki (1994, 1995) and 
Mauche and Raymond (2000) and 
offer an explanation for this behaviour by examining the interplay 
between the thermal and tidal instabilities which affect the accretion 
disks in these systems.

\end{abstract}


\section{Introduction}
 
Cataclysmic variables (CVs) are close binaries in which mass transfer is 
taking place. The primary star is a degenerate white dwarf and the secondary  
a late-type K or M dwarf. The orbital periods are typically measured in 
hours, and in non-magnetic CVs an accretion disk is commonly formed around 
the white dwarf. One group of CVs which undergo quasi-periodic outbursts is 
called the dwarf novae (DN). Outbursts are thought to occur as a result of 
increased accretion onto the white dwarf due to instabilities in the disk. 
For a review of these binaries, see Warner (1995).

SW UMa is a member of the SU UMa class of dwarf novae. These systems have 
orbital 
periods $\leq2$~hr and show at times longer, brighter outbursts called  
superoutbursts. The cause of these superoutbursts has been proposed to be a 
combination of the disk instability plus an additional tidal instability 
(Osaki 1995). SW UMa itself (orbital period $\approx~81$~mins) displays 
outbursts every $\sim450$ days (Ritter and Kolb 1998).

On 1997 October 18th SW UMa was reported to be in superoutburst 
(Mattei et~al. 1997), reaching a peak visual magnitude m$_{\rm v}\approx10.4$ 
at 1997 October 19.969.\footnote{Note that Howell et al. (1995) would only have 
classified the outburst as intermediate since it was neither as bright nor 
as long ($\sim16$~days, see Fig. 1) as previous superoutbursts.}
Coincidentally, $\approx13$~hr later (at 1997 October 20.549) 
SW UMa was serendipitously detected and observed with the $\ro$ $\wfc$. 
At this point, SW UMa was one of the brightest objects in 
the EUV sky. The $\ro$ observations continued over the next $\approx19$~hr.

\begin{figure*}
\vspace{8.0cm}
\includegraphics{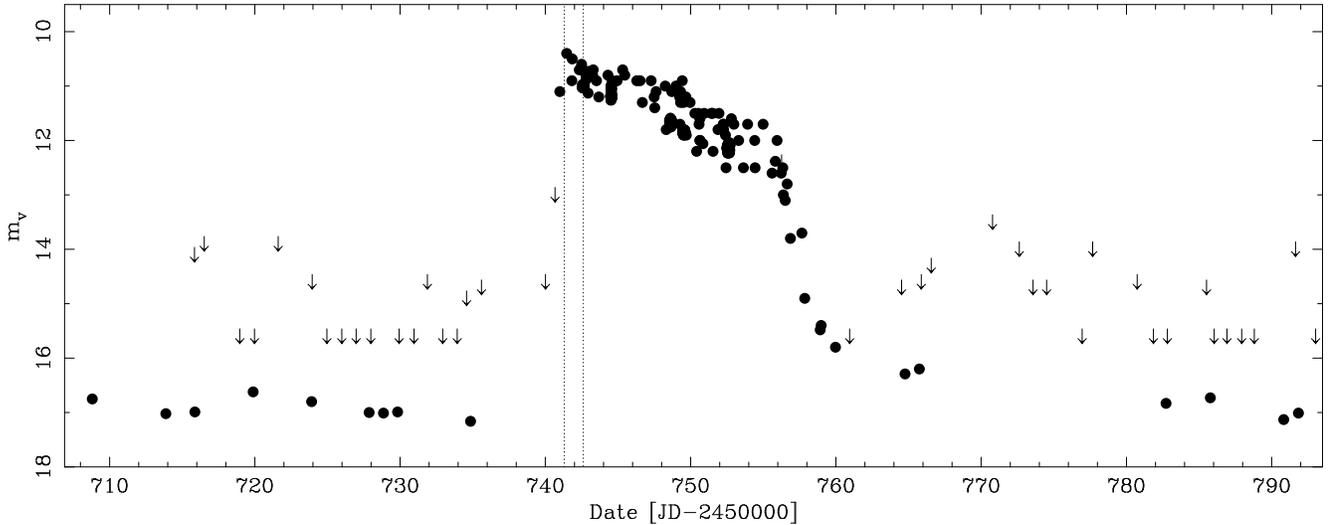}
\caption{Optical light curve of the October 1997 superoutburst of SW UMa, 
from AAVSO records (Mattei 2000). The arrows indicate upper limits on the 
observed magnitude. The dotted vertical lines indicate the extent of the 
simultaneous $\ro$ $\wfc$ observations.}
\end{figure*}

\section{The $\ro$ $\wfc$ observation}

Launched on 1990 June 1, $\ro$ carried two co-aligned instruments, 
an X-ray telescope and the Wide Field Camera 
(Sims et~al. 1990) covering the EUV waveband. After an initial 6-month 
survey phase a programme of pointed observations began, continuing until 
the mission terminated in February 1999. Most observations with the $\wfc$ were 
conducted through one of two broad-band filters, S1 which covered 
the range 60$-$140{\AA} (90$-$200~eV), and S2
which covered 112$-$200{\AA} (60$-$110~eV).

We are currently re-processing and analysing the entire database of $\wfc$ 
pointed phase observations (Pye et~al., in preparation). In the course of this work, 
we discovered a serendipitous off-axis detection of SW~UMa in superoutburst, 
during a pointed observation of the quasar 4C$+$55.16 ($\ro$ observation sequence 
number 703892). The S1 filter was in place throughout the observation, 
which began at 1997 October 20.549. SW UMa was visible near to the edge of  
the detector, 126.5$\arcmin$ off-axis. 
Since the field of view of the $\wfc$ is much greater than that 
of the X-ray instrument in use at the time (the High Resolution Imager), 
it was not seen by that telescope (the $\wfc$ has a field of view 
$\approx5^\circ$ in diameter, the HRI $\approx30\arcmin$). 
The observation lasted for $\approx19$~hr, although SW UMa was not 
observed continuously due to various 
observing constraints (e.g.~earth occultation, passage 
through radiation belts, scheduling of other astronomical targets).

\begin{figure}
\vspace*{14cm}
\includegraphics{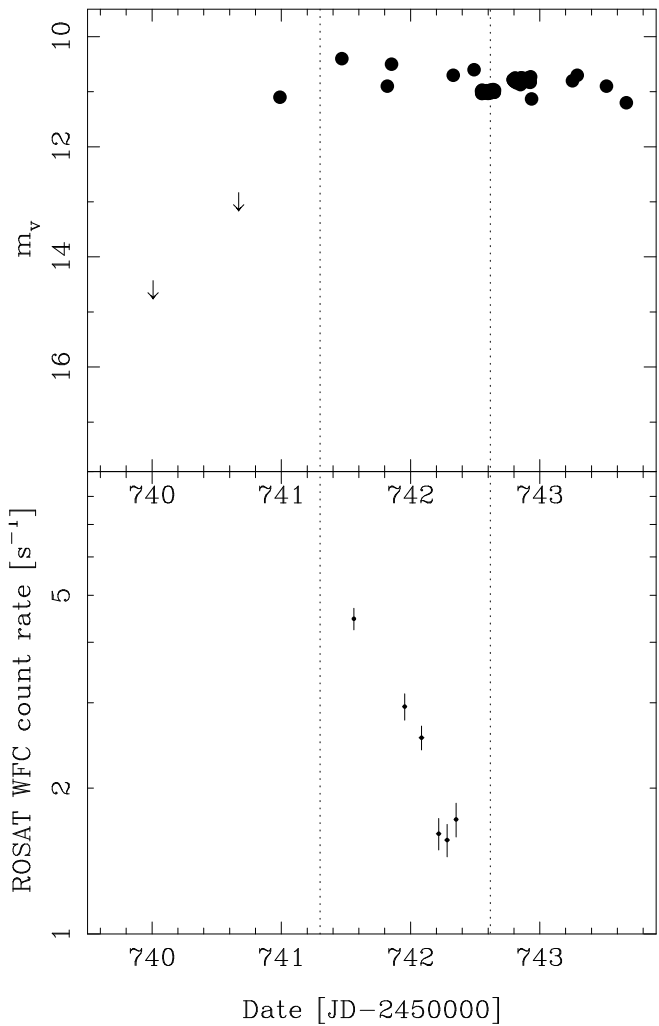}
\caption{Bottom: the EUV light curve obtained by the $\ro$ $\wfc$. The 
corresponding section of the optical light curve is shown above for 
comparison.}
\end{figure}

SW UMa had not previously been detected 
with the $\ro$ $\wfc$ in a pointed phase observation. 
Rosen et~al. (1994) detected the system in quiescence with the 
Position Sensitive Proportional Counter 
($\pspc$) in 1992 April/May, but it was not seen by the $\wfc$ 
at that time.
However, it was marginally ($3.1\sigma$) detected 
during the $\wfc$ all-sky survey, in the S1 filter, when it was also in quiescence. 
Wheatley (1995) measured the survey count rate at  
0.0075 count~s$^{-1}$ (68\% confidence range 0.0044$-$0.011 
count~s$^{-1}$). 
The system has been observed once before in the EUV/soft X-ray regime  
during superoutburst, 
by $\EXOSAT$ during March 1986 (Szkody, Osborne and Hassall 1988). On that  
occasion, it was detected by the low energy detector LE1's 
3000 Lexan filter (0.05$-$2.0 keV) and also in the Al/P filter (0.04$-$0.08, 
0.13$-$1.8 keV), but not by the medium energy ME detector. The 3000 Lexan 
filter count rate increased by a factor 44 above its quiescent value. 

\section{Analysis}

\subsection{The optical and EUV light curves}

Figure 1 shows the 1997 October optical light curve across the 
whole of the superoutburst (from AAVSO records, Mattei 2000). 
The dotted vertical lines 
indicate the position of the $\wfc$ observation, for comparison. 
Note that in the optical the plateau phase lasted for 
$\sim15$~days, during which the brightness declined steadily from a peak 
of m$_{\rm v}\approx10.4$ to m$_{\rm v}\approx13.5$. 
It then rapidly declined back to quiescence. 

The EUV light curve (Figure 2) was constructed using the Starlink software 
Asterix (Allan and Vallance 1995) 
together with the $\wfc$ data reduction package WFCPACK (Denby and 
McGale 1995). The count rates have been corrected for detector degradation 
and are equivalent to ``at launch'' values.\footnote{Over the course of its operational 
lifetime the sensitivity of the $\wfc$ degraded significantly. In 
particular, the spacecraft accidentally pointed near the Sun in 1991 January, 
resulting in a factor $\sim7$ decrease in the detector sensitivity. 
By 1997 October, when these observations of SW UMa were made, the 
sensitivity was just 0.11 of that ``at launch''. A full description of the 
$\wfc$ calibration history will be given by Pye et~al. (in preparation).} 

The source was brightest in the EUV at the start of the $\wfc$ observation,  
at a count rate of $\sim4.5$ count s$^{-1}$, 
$\approx13$~hr after the optical outburst had been first observed. 
At this point, it was the third brightest object in the EUV 
sky between $60-140${\AA}, after the steady white dwarf 
sources HZ43 ($\sim14.5$ count s$^{-1}$) and 
H1504 ($\sim4.8$ count s$^{-1}$). 
It had also brightened by a factor $400-1000$ over the quiescent 
$\wfc$ count rate measured by Wheatley (1995). 
 
Over the next $\approx19$~hr the measured EUV flux 
declined by a factor $\sim2.5$, in contrast to the optical brightness, 
which was essentially static but declined by no more than a 
factor $\sim1.5$. The EUV flux 
then appeared to level out at $\sim1.7$ count s$^{-1}$.

\subsection{A period search on the EUV light curve}

We have conducted a period search on the SW UMa $\wfc$ data, 
in particular to search 
for any evidence of the $\approx81$ minute orbital period, $\approx84$ 
minute ``superhump'' period, and other shorter periods such as a 15.9 minute 
optical period reported by Shafter, Szkody and Thorstensen (1986), 
which may be associated with the rotation period of the white dwarf. 
The data were binned in 30 second intervals for this analysis but 
the period search proved negative. 
A search for faster periods such as 
possible dwarf novae oscillations, using data binned at one second intervals, 
also proved negative.

\section{Discussion}

\subsection{The 1997 EUV superoutburst light curve}

The rapid decline in the EUV luminosity of SW UMa 
so soon after rising to superoutburst is surprising. 
For example, the EUV light curves of the dwarf novae U Gem and SS Cyg 
during normal outbursts follow the behaviour of the 
corresponding optical light curves throughout the optical plateau phase
and during the subsequent rapid decline (e.g~Long et~al. 1996, 
Mauche 1998). 
No rapid decline in the brightness of the EUV light curve is seen 
so early in the outburst. 

However, similar behaviour to that seen here in SW~UMa 
has been reported recently in the superoutburst EUV light curve of the 
related SU UMa-type system OY~Car (Mauche and Raymond 2000). This binary 
was observed for just over three days by the Extreme Ultraviolet Explorer 
($\euve$) 
shortly after it went into superoutburst in March 1997. The Deep Survey 
instrument light curve declined rapidly from a peak brightness of 
$\approx0.21$ count s$^{-1}$ at the start of the observation, 
to $\approx0.07$ count s$^{-1}$ after just $\sim1.5$ days. The optical 
light curve declined much more slowly over the same period. 
Then, perhaps more surprisingly, the EUV brightness 
increased after $\sim2.5$ days 
to $\approx0.1$ count s$^{-1}$ (see Figure 2 of Mauche and Raymond 2000). 
Mauche and Raymond draw comparisons to similar behaviour in 
the Voyager FUV light curve of VW Hyi in superoutburst (Polidan and 
Holberg 1987). An early EUV/soft X-ray dip may also be present 
in that object's corresponding $\EXOSAT$ light curve (Pringle et~al. 1987). 


We now offer an explanation for the 
behaviour of the EUV light curves of SU~UMa-type binaries 
during superoutburst, in terms of the interplay
between the thermal and tidal instabilities which affect the accretion 
discs in these systems. We expand on the ideas put forward by 
Mauche and Raymond (2000) for the behaviour seen in OY Car and VW Hyi, 
and first discussed by Osaki (1994, 1995). 
We assume that the decline in the WFC count rate 
represents a true decrease in the total EUV flux. With observations in a
single filter we cannot rule out the possibility that the observed drop in
count rate is instead due to a spectral variation (e.g. a decrease in
temperature)\footnote{Wheatley et~al. (1996a) and Wheatley et~al. (1996b) 
discuss the possible effects of spectral variations in dwarf novae 
on observed WFC count rates.}, although we consider this unlikely.

The lightcurves of SU~UMa systems show frequent, relatively short and weak
normal outbursts interspersed by long, large amplitude superoutbursts.  
The normal outbursts are thought to arise because of a thermal-viscous
driven disc instability due to the partial ionization of hydrogen (see
Cannizzo 1993 for a review). Each normal outburst only involves the
accretion of a small fraction of the total disc mass onto the white dwarf.
Consequently, there is a gradual accumulation of mass and angular momentum
in the disc.  As a result, the outer disc radius expands until the tidal
resonance radius is reached (in systems with a mass ratio below $\sim
0.33$, a resonance arises because of a 3:1 commensurability between the
disc and binary frequencies, Whitehurst and King 1991). At this point the
enhanced tidal interaction between the disc and the secondary star removes
angular momentum from the outer disc and inhibits further evolution of the
disc radius. Osaki (1995) proposed that the superoutbursts of SU~UMa
systems arise because of this thermal-tidal instability which
spreads through the disc via a heating wave, on the short
thermal timescale. The tidal interaction subsequently
increases the mass accretion rate through the disc, triggering
a superoutburst.  

The observed EUV emission from an SU~UMa star in outburst 
is produced by the hot, inner regions of the accretion disc. 
Once in outburst the surface density of the inner disc,
and hence the resulting EUV emission,
will decay on the timescale $t_{\rm visc} \sim R^2/(\alpha c_s H)$, where
$R$ is the radius within the disc, $\alpha$ is the Shakura-Sunyaev
viscosity parameter, $c_s$ is the local sound
speed and $H$ is the disc semi-thickness (Frank, King and Raine
1992). Typical parameters of the inner disc in outburst
($R \sim 10^9$ cm, $\alpha \sim 0.1$, $c_s \sim 10$
km s$^{-1}$, $H \sim 0.1 R$)  yield $t_{\rm visc} \sim 1$ day. 
However, as outlined above, in a superoutburst the expansion of the
accretion disc causes the tidal radius to be exceeded and the tidal
instability to set in. The EUV emission rate recovers when the inner disc
is resupplied with mass as the tidal instability increases the mass
transfer rate through the disc.
We can estimate the timescale for the growth of the tidal instability from
the delay between the time of super-maximum and the first appearance of
superhumps in the optical light curve (superhumps are small-amplitude
modulations seen in the optical light curves during superoutburst, and are
caused by the accretion disc becoming eccentric due to the action of the
tidal instability). This timescale is between 3$-$7 days in the case of
SW~UMa (see Warner 1995, Table 3.7 and references therein). So, if we
associate the observed EUV emission of SW~UMa with viscous dissipation in
the inner regions of the accretion disc we can understand the decline seen
after super-maximum as the reduction in the surface density of the inner
disc during outburst, with an expected e-folding time of $\sim 1$ day. We
predict a subsequent rise in the EUV flux after $< 3 - 7$ days, as the
inner regions of the disc are re-supplied with mass by the action of the
tidal instability, although none is seen in the data presented here since
the ROSAT WFC observation ended $\sim 1$ day after the start of the
optical superoutburst. The optical light curve remains approximately constant
during this period as the optical emission of the accretion disc is 
dominated by its outer regions, where the surface density
decays on a much longer timescale.

The observations of OY~Car by Mauche and Raymond are also in agreement
with this picture. The EUV light curve is seen to decay over 
1.5 days, close to the viscous timescale of the inner disc, 
and recover after $\sim 2.5$ days; the time between supermaximum and
the first appearance of superhumps in the optical light curve in
this system is $\sim 2 - 3$ days (Warner 1995).

We suggest that the behaviour seen here in the EUV light curve of SW~UMa, 
by Mauche and Raymond (2000) in OY~Car,  and possibly in the $\EXOSAT$ 
soft X-ray light curve of VW~Hyi (Pringle et~al. 1987),  
is a common feature of superoutbursts in 
SU~UMa systems, compared with normal outbursts in U Gem-like dwarf novae. 
The EUV light curves can, therefore, provide both observational 
clues to the superoutburst mechanism and an observational means of 
discriminating between normal outbursts and superoutbursts in these systems.
It is 
unfortunate that the last major mission observing the EUV section of the 
electromagnetic spectrum, $\euve$, has recently been terminated by NASA. No  
new observatories are planned for this waveband in the near future. 

\subsection{Transients observed in the EUV}

At peak brightness, at the start of the $\wfc$ observation, 
SW UMa was the third brightest object in the EUV 
sky between $60-140${\AA}, and the fourth brightest object 
ever seen by the $\wfc$. The brightest was the EUV transient 
RE~J1255$+$266, observed in 1994 June with the S2 filter  
at a peak degradation-corrected count rate of 76.5\,count~s$^{-1}$ 
(Dahlem et~al. 1995). 
Watson et~al. (1996) subsequently showed that RE~J1255$+$266 was a WZ~Sge type 
CV system in outburst (we discuss this outburst in more detail in the 
Appendix). Bright transient events have also been seen by $\euve$. 
For example, U Gem was one of the brightest objects in the EUV sky during 
its 1993 December outburst when it was monitored by $\euve$ 
(Long et~al. 1996). The EUV sky monitor $\ALEXIS$ has also observed a variety of 
transients, including cataclysmic 
variables in outburst, such as the SU~UMa-type binary VW Hyi (Roussel-Dupree and 
Bloch 1996). 
Clearly, then, during outburst 
cataclysmic variable systems can provide some of the most intense sources 
of EUV radiation in the sky. However, we also note that $\ALEXIS$ observed a number 
of shorter ($\la$ 1~day) transients of unknown origin and nature. These events 
could be previously unrecognised CVs or, for example, due to geocoronal stellar flares.

\section*{Acknowledgements}

We acknowledge the support of PPARC for the $\ro$ project in the UK.
This research has made use of data obtained from 
the Leicester Database and Archive Service (LEDAS), operated by the Dept. of Physics 
and Astronomy, Leicester University, UK, and from the SIMBAD database 
operated by CDS, Strasbourg, France. 
In this research, we have used, and acknowledge with
thanks, data from the AAVSO International Database,
based on observations submitted to the AAVSO by variable
star observers worldwide.

{}

\large

\noindent
{\bf APPENDIX: The EUV transient RE~J1255$+$266 -- a normal or superoutburst of 
a WZ~Sge system?}

\normalsize

The exceptionally bright EUV transient source RE~J1255$+$266 was discovered 
by the $\ro$ $\wfc$ in 1994 June during rapid exponential decline 
(Dahlem et~al. 1995). The $\wfc$ count rate decreased exponentially over a 
period of four days, with an e-folding time of about one day. The rise was 
not covered, and the source was not detected in another $\wfc$ observation 
eight days later, consistent with a continuing exponential decline. 
The outburst was not observed optically. 
Watson et~al. (1996) later identified the optical counterpart spectroscopically 
and concluded that it was probably a WZ~Sge-type CV system that had been in 
superoutburst (WZ~Sge systems, a subset of the SU~UMa's,  
appear to show a strong preference for 
superoutbursts that last weeks, rather than normal outbursts which last for 
a few days). Recently, though, Wheatley, Burleigh and Watson (2000) have shown 
that the object was in quiescence just four days before the
discovery observation, and concluded that this was most likely only a
normal outburst of a WZ~Sge system. The superoutburst EUV light curve of 
SW~UMa (Figure 2), together with the $\euve$ observation of OY~Car discussed 
by Mauche and Raymond (2000), now raises the possibility that the rapid EUV 
decline of RE~J1255$+$266 
actually occured at the beginning of a much longer optical superoutburst.
However, both SW~UMa and OY~Car show an EUV decline for $\la1$~day, 
which then flattens off and possibly rises again. In RE\,J1255$+$266 the
decline continued at a constant exponential rate for at least 4\,days and
probably for more than 12\,days. We take this as further evidence that the
outburst of RE~J1255+266 detected by $\ro$ was probably a
normal outburst of a WZ~Sge CV system, not a superoutburst.

\end{document}